\documentclass[12pt,preprint]{aastex}
\def\paren#1{\left( #1 \right)}

\begin{document}
\title{GRB 021004: Reverse Shock Emission}
\author{Shiho Kobayashi$^{1,2}$ and Bing Zhang$^{2}$}
\affil{
$^{1}$Dept. of Physics,
Pennsylvania State University, University Park, PA 16802\\
$^{2}$Dept. of Astronomy \& Astrophysics 
Pennsylvania State University, University Park, PA 16802}
\begin{abstract}
 We show that the re-brightening in the GRB 021004 optical afterglow
 light curve can be explained within the framework of the standard
 fireball model. The optical light curve of the forward shock emission
 is expected to rise initially and to decay after the typical
 synchrotron frequency crosses the optical band. We show that such a
 rising phase emission was caught for GRB 021004, together with a
 reverse shock emission. With the standard values of parameters obtained
 in other afterglow observations, we can construct example cases in
 which theoretical estimates reasonably fit the broadband
 observations. Therefore, the early re-brightening might be a common
 feature in optical afterglows.  
\end{abstract}
\maketitle
\section{introduction}

GRB 021004 triggered HETE II on 2002 October 4 at 12:06:13 UT. The burst
lasted $T\sim 100$ seconds (Shirasaki et al. 2002) and
the $7-400$ keV fluence  was $\sim 3.2\times10^{-6}$ ergs cm$^{-2}$ 
(Lamb et al. 2002). The prompt localization of GRB 021004 by HETE II 
allowed the follow-up of the afterglow at very early time. Fox (2002)
detected an optical transient $\sim 9$ mins after the trigger at the
level of $R\sim 15.56$ mag. 
Torii, Kato and Yamaoka  (2002) observed the error box of the
burst  $\sim 3.5$ mins after the trigger, yielding upper limit around
$R\sim 13.6$ mag at the position of the transient discovered by Fox (2002). 
The spectroscopic observations of the optical afterglow revealed an
emission line interpreted as Ly-$\alpha$ at $z=2.328$ (Mirabal et al. 2002).
Assuming $\Omega_0=0.3,\lambda_0=0.7$ and $h=0.6$, the isotropic
gamma-ray energy is $\sim 5.6\times10^{52}$ ergs. 

The dense sampling of the afterglow light curve at early time revealed
the peculiarity, a major re-brightening around $\sim 0.1$ day after the
trigger and a short time scale variability around $\sim 1$ day.
The slope of the afterglow is $\sim t^{-0.7}$ for the earliest
three observations (9-17 min after trigger) by Fox (2002),
and after the luminosity increases around $\sim 0.1$ days,
it decayed with  $\sim t^{-1.05}$ as  usual afterglows do.
These temporal features might be modeled by refreshed shocks (Zhang and
M\'{e}sz\'{a}ros 2002). Recently, Lazzati et al. (2002) interpreted
the features as due to enhancements in the ambient medium density.
In this Letter, we show that the major re-brightening around $\sim 0.1$ day
can be explained within the standard fireball model. We propose
that the reverse shock emission dominated the optical band at
early times, leading to the observed peculiarity.

\section{The model}
Consider a relativistic shell (fireball ejecta) with
energy $E$, a  Lorentz factor $\eta$ and a width $\Delta_0$ expanding
into the homogeneous ISM with a particle number density $n$. When the
shell sweeps a large volume of the ISM, it is decelerated and
the kinetic energy is transfered to the ISM by shocks (e.g. Kobayashi, Piran
\& Sari 1999). The shocked ISM forms a relativistic blast wave (forward
shock) and emits the internal energy via synchrotron process.

The emission from a reverse shock was also predicted (M\'{e}szaros \&
Rees 1997; Sari \& Piran 1999a). When the reverse shock crosses the shell,
the forward shocked ISM and the reverse shocked shell carry comparable
amounts of energy. However, the typical temperature of the shocked shell
is lower since the mass density of the shell is higher. Consequently,
the typical frequency in the shocked shell is lower. A prompt
optical emission from GRB~990123 (Akerlof et al. 1999) can be regarded
as this emission (Sari \& Piran 1999b; Kobayashi \& Sari 2000; 
Soderberg \& Ramirez-Ruiz 2002).

\subsection{Forward Shock}
Observations of optical afterglows usually start around several hours after
the burst trigger. Since, at such a late time, the typical synchrotron
frequency of the forward shock emission $\nu_{m,f}$ is lower than
optical band $\nu_R \sim 5\times 10^{14}$Hz, 
the evolution of afterglows is well described by a single power law,
except for the jet break (Rhoads 1999; Sari, Piran \& Halpern 1999).
However, an optical light curve is expected to peak at early time 
when the typical synchrotron frequency crosses the optical band.
Before the peak time $t_{m,f}$, the  luminosity increases as $\propto
t^{1/2}$, and reaches to the maximum flux $F_{\nu,max,f}$, and then
decays as $\propto t^{3(1-p)/4}$ where $p$ is the index of the power law
distribution of random electrons accelerated at shock. 
Using the results in Sari, Piran and Narayan (1998), we get
\begin{eqnarray}
\nu_{m,f}(t) &\sim& 5.1 \times 10^{15} (1+z)^{1/2} \epsilon_B^{1/2}
\epsilon_e^2 g^2 E_{52}^{1/2} t_d^{-3/2} ~\mbox{Hz},\\
\nu_{c,f}(t) &\sim& 2.7\times10^{12}  (1+z)^{-1/2}
\epsilon_B^{-3/2} E_{52}^{-1/2} n_0^{-1} t_d^{-1/2}
~\mbox{Hz},\\
t_{m,f}&\sim& 2.9
(1+z)^{1/3} \epsilon_B^{1/3} \epsilon_e^{4/3} g^{4/3}E_{52}^{1/3} 
\nu_{R,15}^{-2/3} \mbox{days},
\label{eq:tm}\\
F_{\nu,max,f}&\sim& 1.1\times10^2
a_\nu (1+z) \epsilon_B^{1/2} E_{52} n_0^{1/2} D_{28}^{-2}
~\mbox{mJy}.
\end{eqnarray}
where  $\nu_{c,f}$ is the cooling frequency, $\epsilon_B$ and
$\epsilon_e$ are the fractions of the shock energy given to magnetic
field and electrons at the shock, $g=(p-2)/(p-1)$, $E_{52}=E/10^{52}$
ergs, $n_0=n/1$ proton cm$^{-3}$, $\nu_{R,15}=\nu_R/10^{15}$Hz, and here
$t_d$ is the observer's time in unit of day, $D_{28}$ is the luminosity
distance in unit of $10^{28}$ cm, $a_\nu$ is a correction factor to the
extinction along the line of sight to the burst. 

Assuming that the re-brightening with the peak luminosity of $\sim 1$ mJy
around $\sim 0.1$ day is caused by this peak, we get the following formulae
for $\epsilon_e$ and $n_0$ as functions of $\epsilon_B$ and other known
parameters. 
\begin{eqnarray}
\epsilon_e  &\sim& 9.4 \times10^{-2}
\epsilon_B^{-1/4}
\paren{\frac{1+z}{3.3}}^{-1/4}
\paren{\frac{g}{0.29}}^{-1}
\paren{\frac{t_{m,f}}{0.1 \mbox{day}}}^{3/4}
\paren{\frac{E_{52}}{5.6}}^{-1/4}
\paren{\frac{\nu_{R,15}}{0.5}}^{1/2},
\label{eq:epse}
\\
n_0 &\sim&8.0\times10^{-4}
\epsilon_B^{-1}
\paren{\frac{a_\nu}{0.8}}^{-2}
\paren{\frac{1+z}{3.3}}^{-2}
\paren{\frac{F_{\nu,max,f}}{1 \mbox{mJy}}}^{2}
\paren{\frac{E_{52}}{5.6}}^{-2}
\paren{\frac{D_{28}}{6.8}}^{4}.
\label{eq:n}
\end{eqnarray}
Holland et al. (2002a) claim the extinction $A_V=0.26$ mag, which 
implies a $\sim 20 \%$ correction in the R-band. The slope $\sim -1.05$
of the optical afterglow at late time implies $p\sim2.4$. However,
there is still some debate about the value of $p$ (Sako \& Harrison
2002; Holland et al. 2002a; Pandey et al. 2002). We will discuss
two cases $p=2.2$ and $2.4$.

\subsection{Reverse Shock}
The evolution of reverse shocks is classified into two cases
(Kobayashi 2000). If the initial Lorentz factor of the shell
$\eta$ is larger than a critical value
$\eta_c= (3E/32\pi n m_p c^2 \Delta_0^3)^{1/8}$ where $m_p$ is the mass of
proton, the reverse shock becomes relativistic in the frame of unshocked
shell material during crossing the shell, and drastically decelerates
the shell (thick shell case). If $\eta< \eta_c$, the reverse shock can
not decelerate the shell effectively (thin shell case). 
According to the internal shock model, the initial width of the shell
$\Delta_0$ is given by the intrinsic duration of the GRB $\sim cT/(1+z)$ 
(Kobayashi, Piran \& Sari 1997).
\begin{equation}
\eta_c \sim 190
     ~n_0^{-1/8}
     \paren{\frac{1+z}{3.3}}^{3/8}
     \paren{\frac{T}{100 \mbox{sec}}}^{-3/8}
     \paren{\frac{E_{52}}{5.6}}^{1/8}.
\label{eq:etac}
\end{equation}
The Lorentz factor at the shock crossing time is given given by  
$\gamma_\times \sim \min[\eta, \eta_c]$. The shock crossing time is 
$t_\times \sim (\gamma_\times/\eta_c)^{-8/3}T$
(Sari \& Piran 1999a; Kobayashi 2000).

Since at the shock crossing time, the forward and reverse shocked regions
have the same Lorentz factor  and internal energy density $e$, the
cooling frequency of the reverse shock $\nu_{c,r}$ is equal to that of
the forward shock $\nu_{c,f}$.
\begin{equation}
\nu_{c,r}(t_\times) \sim \nu_{c,f}(t_\times).
\end{equation}
The typical frequency of synchrotron emission is proportional to
the electron's random Lorentz factor squared and to the magnetic
field and Lorentz boost. The Lorentz boost and the magnetic field
$\propto \sqrt{e}$ are the same for the two shocked regions, while
the random Lorentz factor is proportional to $\bar{\gamma}_\times$
in the reverse shocked region and to $\gamma_\times$ in the forward shocked
region where $\bar{\gamma}_\times$ is the Lorentz factor of the shocked
shell material in the frame of the unshocked shell material. 
Using a relation $\gamma_\times\bar{\gamma}_\times\sim\eta$,
the reverse shock frequency at the crossing
time (the peak time) is given by (Sari \& Piran 1999a)
\begin{equation}
\nu_{m,r}(t_\times) \sim
\frac{\eta^2}{\gamma_\times^4} ~\nu_{m,f}(t_\times).
\label{eq:numr}
\end{equation}
The peak flux at the typical frequency is proportional to the number
of electrons and to the magnetic field and the Lorentz boost. From the
energy conservation, the mass of the shell is larger by a factor of
$\gamma_\times^2/\eta$ at the crossing time than that of the ISM
swept by the forward shock. Since the number of electrons is proportional to
the mass, we get
\begin{equation}
F_{\nu,max,r}(t_\times) \sim \frac{\gamma_\times^2}{\eta} F_{\nu,max,f}
\end{equation}

Even though the hydrodynamic evolution of ``thin'' and ``thick''
shells are very different, the time dependences of the emission
are similar (Kobayashi \& Sari 2000). If the optical band $\nu_R$ is
below the typical synchrotron frequency $\nu_{m,r}$, the luminosity
decays as $\sim t^{-0.5}$. If $\nu_R > \nu_{m,r}$, it decreases 
as $\sim t^{-2}$ (Kobayashi 2000).

First we assume that the optical band is below the typical frequency 
of the reverse shock emission at the shock crossing time 
$\nu_{m,r}(t_\times) > \nu_R \sim \nu_{m,f}(t_{m,f})$.
Using $\nu_{m,f} \propto t^{-3/2}$ (Sari, Piran \& Narayan 1998) and 
eq. (\ref{eq:numr}), we get
$\eta > (T/t_{m,f} )^{3/4} \eta_c^2 \sim 0.03 \eta_c^2 > 1000$
for the standard values of parameters assumed  in eq. (\ref{eq:etac}).
Then, GRB021004 should be a thick shell case. 
However, in a thick shell case, we can show 
$F(t_{eo})/F_{\nu,max,f} < T t_{m,f}^{3/4}/t_{eo}^{7/4} \sim 0.6$
where $F(t_{eo})$ is the the optical flux from the reverse shock at the
earliest observations by Fox (2002) at  $t_{eo}\sim 0.01$ days. 
This limit is inconsistent with the observations,
and hence $\nu_{m,r}(t_\times) < \nu_R$.

If $\nu_{m,r}(t_\times) < \nu_R$, the reverse shock emission simply
decreases as $t^{-2}$. Since the extrapolation of the light curve from
the data points by Fox (2002) to the earlier time $T \sim 10^{-3}$
day with a power law of $t^{-2}$ violates the upper-limits by 
Torii et al (2002) (see fig1), GRB 021004 should be a thin shell case,
in which the peak of the reverse shock emission is delayed, i.e. 
$t_\times > T$.

Using eqs. (\ref{eq:epse}) and (\ref{eq:n}) with 
$p=2.4$, $t_{m,f}\sim 0.06$ days, $F_{\nu,max,f}\sim 1.3$ mJy (R-band),
$E=5.6\times10^{52}$ergs, and $z=2.3$,
we search for a set of parameters $(\epsilon_B,\eta)$ with which
the theoretical estimates give a reasonable fit to all observations. With
a normalized Lorentz factor $\kappa\equiv\eta/\eta_c$,
we can show that the crossing time $t_\times \propto \kappa^{-8/3}$
and the optical flux from the reverse shock at the crossing time
$F(t_\times)= (\nu_R/\nu_{m,r})^{-(p-1)/2} F_{\nu,max,r}
\propto \kappa^p \epsilon_B^{(2-p)/8}$. Since the optical flux 
$F(t_\times)$ depends on $\epsilon_B$ very weakly 
$\propto \epsilon_B^{-1/20} (p=2.4)$ and $\epsilon_B^{-1/40} (p=2.2)$, 
$\epsilon_B$ is not well determined from the light curve,
or equivalently, we can explain the peculiar behavior of the light curve
with a wide range of $\epsilon_B$. When changing the value of $\kappa$,
the position of the peak moves along a line of $F \propto t^{-3p/8}$ on
the flux - time plane, and $\kappa \sim 0.55$ gives the best fit to the
observations.  When we fit  the observations at $>0.1$ day
with a  flatter power law of $t^{3(1-p)/4}=t^{-0.9} (p=2.2)$, we need to
chose a earlier peak time $t_{m,f} $. Assuming the same
parameters with the $p=2.4$ case, except $t_{m,f}\sim 0.035$ days, the
best fit is given by $\kappa \sim 0.65$. 

Since $\epsilon_e$ and $n$ are determined by eqs. (\ref{eq:epse}) and 
(\ref{eq:n}), $\nu_{m,f}$ and $F_{\nu,max,f}$ do not depend on
$\epsilon_B$. The value of $\epsilon_B$ is constrained only by the
the cooling break. The cooling frequency  $\nu_{c,f} \propto t^{-1/2}$
crosses the optical band at  
$t \sim 0.79 ~\epsilon_B^{-1}
\paren{\frac{1+z}{3.3}}^{3}
\paren{\frac{E_{52}}{5.6}}^3
\paren{\frac{a_\nu}{0.8}}^4
\paren{\frac{F_{\nu,max,f}}{1.3 mJy}}^{-4}
\paren{\frac{\nu_{R,15}}{0.5}}^{-2}
\paren{\frac{D_{28}}{6.8}}^{-8}$
days.
Holland et al. (2002a) reported that there is no evidence for color
evolution between 8.5 hours and 5.5 days after the burst (see, however, 
Matheson et al. 2002; Bersier et al. 2002), and hence
$\epsilon_B<0.14$. A lower limit $\epsilon_B > 8 \times 10^{-5}$ is
required from $\epsilon_e < 1$.

\begin{figure}
\plotone{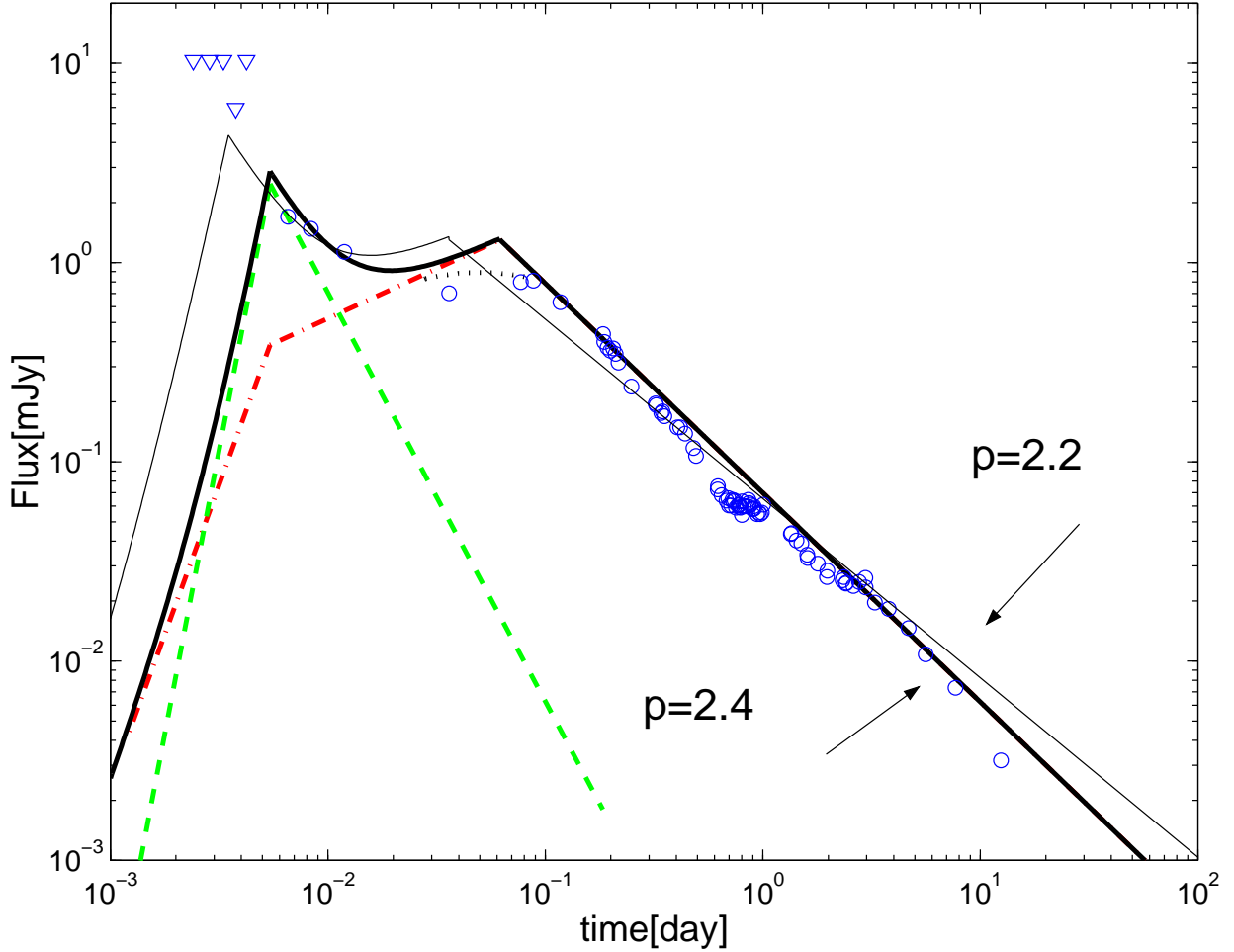}
\caption{Optical light curve: 
(1) modeling with $p=2.4$,
forward shock emission (dashed dotted),
reverse shock emission (dashed), 
total flux (thick solid), 
accurate spectrum mode (dotted).
$\epsilon_B=3.0\times10^{-3}$,
$\epsilon_e=0.28$. 
$n=0.45$ protons cm$^{-3}$, 
$\eta=120$ and
$E=5.6\times 10^{52}$ ergs. 
(2) modeling with $p=2.2$, 
total flux (thin solid). 
$\epsilon_B=3.0\times10^{-3}$,
$\epsilon_e=0.32$. 
$n=0.45$ protons cm$^{-3}$, 
$\eta=140$ and
$E=5.6\times 10^{52}$ ergs. 
Measurements
(circles) and upper-limits (triangles). 
Data from: Bersier et al. 2002b; Fox 2002; Halpern
et al. 2002a,b; Holland et al. 2002c,d; Malesani et al. 2002a,b; Masetti
et al. 2002; Matsumoto et al. 2002a,b; Mirabal et al. 2002a,b;
Oksanen\&Aho 2002; Oksanen et al. 2002; Sahu et al. 2002; Stanek et
al. 2002; Stefanon
et al. 2002; Torii et al. 2002; Uemura et al. 1566; Weidinger et
al. 2002; Winn et al. 2002; Zharikov et al. 2002, following the
calibration of Henden 2002.
 \label{fig1}}
 \end{figure}

Fig \ref{fig1} shows an  example case in which $p=2.4$ and
$\epsilon_B = 3.0\times10^{-3}$ are assumed. This choice leads to
$\epsilon_e \sim 0.28$, $n\sim 0.45$ proton cm$^{-3}$ and $\eta \sim 120$.
These are surprisingly typical values obtained in other afterglow
observations (Panaitescu and Kumar 2002). The dashed and dashed dotted
lines show the optical light curve of the reverse and forward shock 
emission, respectively. The thick solid lines depicts the total flux. 
Around the peak time of the forward shock $\sim 0.1$ day, our estimate
slightly deviates  from the observations. However, in our estimate, we
assumed a simplified synchrotron spectrum which is described by a broken
power law. Since a more realistic synchrotron spectrum is rounded at the
break frequencies (Granot, Piran \& Sari 1999), 
a light curve should be also rounded at the break time (dotted line).
The short time scale variability, which is prominent around $\sim 1$
day, might be produced by ISM turbulence (Wang \& Loeb 2000; 
Holland et al. 2002b; Lazzati et al 2002). The latest data point in Fig 
\ref{fig1} is lower than the extrapolation with a scaling of
$t^{-1.05}$. This might be a signature of the jet break.

Assuming $p=2.2$ and $\epsilon_B = 3.0\times10^{-3}$, we get
$\epsilon_e \sim 0.32$, $n\sim 0.45$ proton cm$^{-3}$ and 
$\eta \sim 140$.  The total flux is shown by the thin solid line.
In this case, we need to assume larger ISM turbulence.

\section{X-ray and Radio Afterglow}
In this section, we assume the values of parameters with which we have
shown the example cases in the previous section, and estimate the X-ray 
and radio afterglows. The extinction correction $a_\nu$ is unity for 
X-ray and radio afterglows, and hence  $F_{\nu,max,f} \sim 1.6$ mJy 
(radio and X-ray).

X-ray Afterglow: Since the X-ray band $\sim 5$keV 
is well above the typical frequency of the reverse shock
emission, the contribution from the reverse shock to the X-ray band is
negligible. The X-ray afterglow should be described only by the forward
shock emission.  The luminosity in X-ray band should decrease as
$t^{(2-3p)/4} \sim t^{-1.3}(p=2.4)$ or $t^{-1.15}(p=2.2)$.
The Chandra X-ray observatory observed the afterglow for a total
exposure of 87ksec, beginning at Oct 5 8:55 UT (Sako and Harrison
2002). The count rate decrease  with a power law slope of $-1.0 \pm
0.2$. The mean 2-10 keV X-ray flux is $\sim 4.3 \times10^{-13}$ ergs
cm$^{-2}$ s$^{-1}$. We estimate  the 5 keV flux at the observational
mean time 1.36 days $\sim 3.1\times10^{-13}$ ergs cm$^{-2}$ s$^{-1}$ for
$p=2.4$ and $\sim 6.4\times10^{-13}$ ergs cm$^{-2}$ s$^{-1}$ for $p=2.2$. 
Our estimates are in a good agreement with the observations.

Radio Afterglow: The forward shock emission in radio band $\sim 10$ GHz
increases as $t^{1/2}$ until the flux reaches to the
maximum $\sim 1.6$ mJy at $\sim 80$ days for $p=2.4$ (dashed dotted
line in fig2) or at $\sim 50$ days for $p=2.2$. After the typical 
frequency $\nu_{m,r}$ crosses the radio band, the reverse shock emission
decays as $\sim t^{-2}$ (dashed line for $p=2.4$). At low frequencies
and early times, self 
absorption takes an important role and significantly reduces the flux. A
simple estimate of the maximal flux (dotted line for $p=2.4$) is the 
emission from the black body with the reverse shock temperature 
(Kobayashi \& Sari 2000). The thick and thin solid 
line depicts the total flux for $p=2.4$ and for $p=2.2$, respectively.
Since the observations (circles) are done in various frequencies, we
scaled the observed value to the expected value at $10$GHz by using a
spectral slope of $1$   
\footnote{
Berger et al. (2002) reported an unusual spectral spectral slope
$F_\nu \sim \nu$ between 8.5 GHz and 86GHz from the observations with
the VLA on October 10.17 UT, and claimed that the spectrum is not due to
a transition from optically-thick ($\nu^2$) to optically thin
($\nu^{1/3}$) emission. The superposition of the forward and reverse
shock emission could give even flatter spectrum. Non-standard emission
mechanism may be needed to explain this unusual spectrum.}.
This burst might also cause a bright radio flare $\sim 1$ mJy around
$\sim 0.5$ day as observed in GRB 990123. When we fit the optical
observations with $p=2.2$, an earlier optical peak time $t_{m,f}$ is 
required. Since the peak time at the radio band is proportional to
$t_{m,f}$, the modeling with $p=2.2$ gives a better fit to the radio
observations. 

\section{Conclusions}
The low-frequency (e.g. optical, IR \& radio) lightcurve from the
forward shock is expected to rise initially and to decay after the
typical synchrotron frequency crosses the observational band. 
Although such a behavior has been observed in the radio band, previous
optical afterglow observations were only made at too late times to
catch the rising phase. However, the swift localization
of GRB 021004 by HETE II allowed the follow-up of the afterglow at very
early time. This burst has so far the earliest detected optical afterglow.
We have shown that the rising phase of the optical emission might be caught 
for the first time in GRB 021004, together with a reverse shock emission.  
The superposition of both the forward shock and the reverse shock
emissions can well fit for the 0.1 day re-brightening feature.
With the standard values of parameters inferred from other afterglow
observations, we have constructed example cases in which theoretical
estimates fit the broadband observations.  We therefore suggest that the
early re-brightening might be a common feature in optical
afterglows. 

The reverse shock emission in GRB 990123 peaked at about $\sim 1$ Jy in
R-band, while the peak in GRB 021004 was only a few mJy. Since GRB
990123 is a very bright burst with a fluence about 100 times that of GRB
021004, we expect GRB 990123 could produce the much brighter reverse
shock emission. Another difference is the typical frequency 
$\nu_{m,r}(t_\times)$. In GRB 990123 it is close to
the R-band (Sari \& Piran 199b), while in GRB 021004 it is estimated as 
$\sim 1.4 \times 10^{12}$ Hz $(p=2.4)$ or
$\sim 8.8 \times 10^{11}$ Hz $(p=2.2)$.
The lower $\nu_{m,r}(t_\times)$
also makes the reverse shock emission dimmer in the R-band
(Kobayashi 2000).

We thank Peter M\'{e}sz\'{a}ros for valuable comments and
the anonymous referee for valuable suggestions.
We acknowledge support through the Center for Gravitational Wave
Physics, which is funded by NSF under cooperative agreement PHY
01-14375, and through NASA NAG5-9192.

\noindent {\bf References}\newline
Akerlof,C.W. et al. 1999, Nature, 398, 400.\newline
Berger,E. et al. 2002, GCN 1613.\newline
Bersier,D.
2002a, submitted to ApJL, astro-ph/0211130.\newline
Bersier,D., Winn,J., Stanek,K.Z. \& Garnavich,P. 
2002b, GCN 1586.\newline
Fox,D.W. 2002, GCN 1564.\newline
Frail,D. \& Berger,E. 2002, GCN 1574.\newline
Granot,J., Piran,T. \& Sari,R. 1999, ApJ, 513, 679.\newline
Halpern,J.P., Armstrong,E.K., Espaillat,C.C. \& Kemp,J. 2002a, GCN 1578.\newline
Halpern,J.P., Mirabal,N., Armstrong,E.K., Espaillat,C.C. \& Kemp,J.
2002b, GCN 1593.\newline
Henden,A. 2002, GCN 1583.\newline
Holland,S.T. et al. 2002a, submitted to AJ, astro-ph/0211094.\newline
Holland,S.T. et al. 2002b, AJ, 124, 639.\newline
Holland,S.T.,Fynbo,J.P.U.,Weidinger,M.,Egholm,M.P.\& Levan,A. 
2002c, GCN 1585.\newline
Holland,S.T.,Fynbo,J.P.U.,Weidinger,M.,Egholm,M.P.,Levan,A.\&Pedersen,H. 
2002d, GCN 1597.\newline
Kobayashi,S 2000, ApJ, 545, 807.\newline
Kobayashi,S, Piran,T \& Sari,R. 1997, ApJ, 490, 92.\newline
Kobayashi,S, Piran,T \& Sari,R. 1999, ApJ, 513, 669.\newline
Kobayashi,S \& Sari,R. 2000, ApJ, 542, 819.\newline
Lamb,D. et al. 2002, GCN 1600.\newline
Lazzati,D., Rossi,E.,Covino,S.,Ghisellini,G \& Malesani,D. 
2002, A\&A in press, astro-ph/0210333.\newline
Malesani,D. et al. 2002a, GCN 1607.\newline
Malesani,D. et al. 2002b, GCN 1645.\newline
Matheson,T et al. 2002, submitted to ApJL, astro-ph/0210403.\newline
Masetti,N. et al. 2002, GCN 1603.\newline
Matsumoto,
K.,Kawabata,T.,Ayani,K.,Urata,Y.\& Yamaoka,H.
2002a, GCN 1567.\newline
Matsumoto,K. 
K.,Kawabata,T.,Ayani,K.,Urata,Y.,Yamaoka,H.\&Kawai,N.
2002b, GCN 1594.\newline
M\'{e}sz\'{a}ros,P. \& Rees,M.J. 1997, ApJ, 476, 231.\newline
Mirabal,J.,Armstrong,E.K.,Halpern,J.P.\& Kemp,J.
2002a, GCN 1602.\newline
Mirabal,J.,Halpern,J.P.,Chornock,R.\&Filippenko,A.V.
2002b, GCN 1618.\newline
Oksanen,A. \& Aho,M. 2002, GCN 1570.\newline
Oksanen,A.,Aho,M,Rivich,K.,Rivich,K.,West,D.\&During,D. 2002, GCN 1591.\newline
Panaitescu,A. \& Kumar,P. 2002, ApJ, 571, 779.\newline
Pandey et al. 2002, submitted to BASI, astro-ph/0211108.\newline
Pooley,G. 2002a, GCN 1575.\newline
Pooley,G. 2002b, GCN 1588.\newline
Pooley,G. 2002c, GCN 1604.\newline
Rhoads,J.E. 1999, ApJ, 525, 737.\newline
Sahu,D.K.,Bhatt,B.C.,Anupama,G.C.\&Prabhu,T.P. 2002, GCN 1587.\newline
Sako,M \& Harrison, F.A. 2002, GCN 1624.\newline
Sari,R. \& Piran,T 1999a, ApJ, 520, 641.\newline
Sari,R. \& Piran,T 1999b, ApJ, 517, L109.\newline
Sari,R., Piran,T \& Halpern,J.P. 1999, ApJ, 519, L17.\newline
Sari,R., Piran,T \& Narayan,R. 1998, ApJ, 497, L17.\newline
Shirasaki,Y. et al. 2002, GCN 1565.\newline
Soderberg,A.M. \& Ramirez-Ruiz,E. 2002, MNRAS, 330, L24.\newline
Stanek,K.Z.,Bersier,D.,Winn,J.\& Garnavich,P. 2002, GCN 1598.\newline
Stefanon,M. et al. 2002, GCN 1623.\newline
Torii,K., Kato,T. \&  Yamaoka,H. 2002, GCN 1589.\newline
Uemura,M.,Ishioka,R.,Kato,T.\&Yamaoka,H. 2002, GCN 1566.\newline
Wang,X. \& Loeb,A  2000, ApJ, 535, 788.\newline
Weidinger,M. et al. 2002, GCN 1573.\newline
Winn,J.,Bersier,D.,Stanek,K.Z.,Garvanich,P.\&Walker,A. 2002, GCN 1576.\newline
Zhang,B.\&  M\'{e}sz\'{a}ros,P. 2002, ApJ, 566, 712.\newline
Zharikov,S.,Vazquez,R.,Benitez,G.\&del Rio,S. 2002, GCN 1577.\newline

 \begin{figure}
\plotone{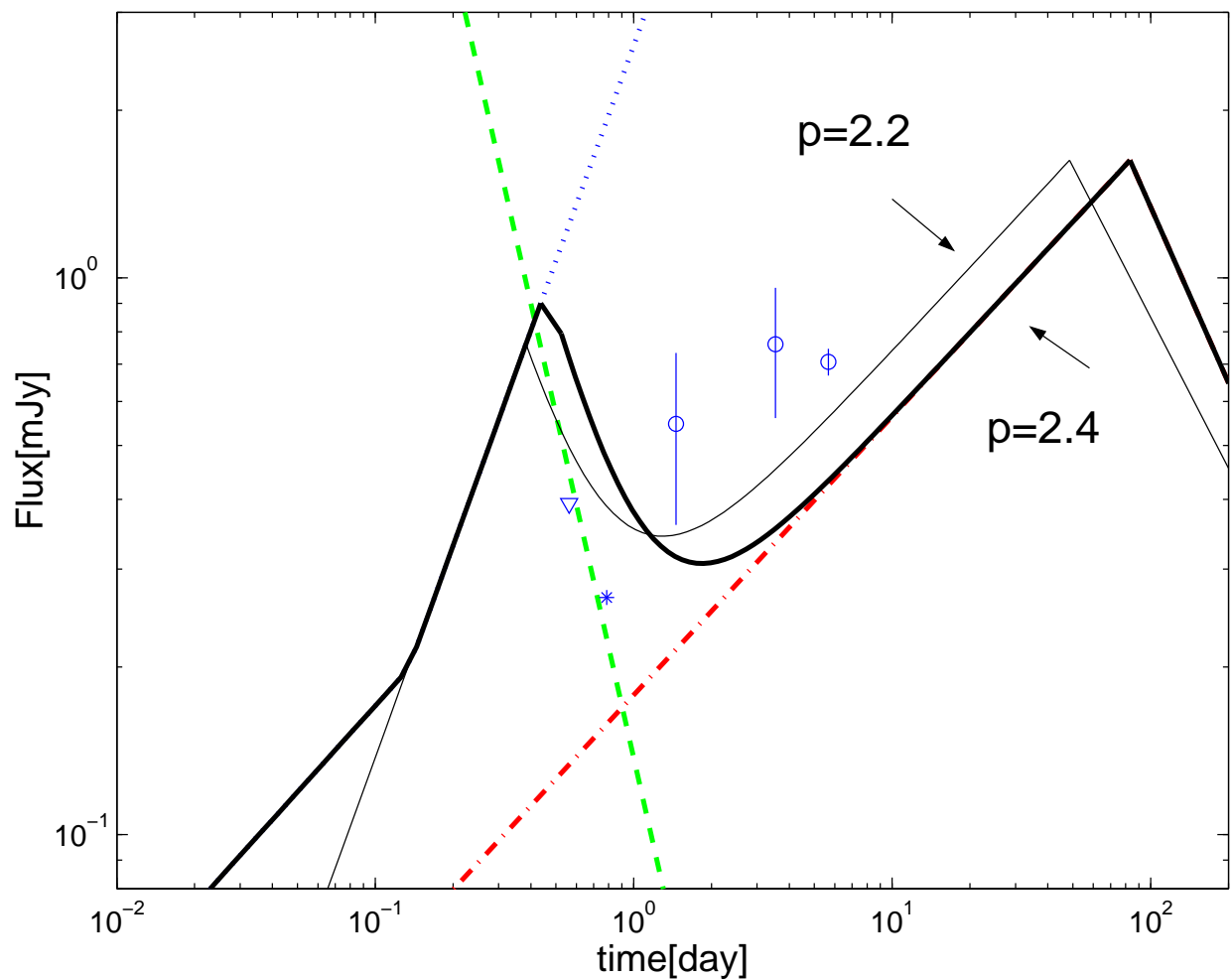}
\caption{
Radio light curve: 
(1) modeling with $p=2.4$,
forward shock emission (dashed dotted), 
reverse shock emission (dashed), 
self-absorption limit (dotted), 
total flux (thick solid). 
(2) modeling with $p=2.2$,
total flux (thin solid). 
The parameters are the same as in Fig.1.
Measurements with error 
bars (circles), measurement without error bar (star) and
upper-limit (triangle). 
Data from: Frail \& Berger 2002; Pooley 2002a,b,c; Berger et
al. 2002b.
 \label{fig2}}
 \end{figure}
\end{document}